\documentclass[aps,prl,preprint,superscriptaddress,showpacs]{revtex4}
\usepackage{epsfig}
\usepackage{subfigure}
\usepackage{amssymb}
\usepackage{graphicx}
\usepackage{amsmath}
\usepackage{natbib}
\newcommand{\mb}{\boldsymbol}
\begin{document}

\title{Measurement of electron-hole friction in an n-doped GaAs/AlGaAs quantum well using optical transient grating spectroscopy}

\author{Luyi Yang}\affiliation{Department of Physics, University of California,
Berkeley, California 94720, USA.} \affiliation{Materials Science
Division, Lawrence Berkeley National Laboratory, Berkeley,
California 94720, USA.}
\author{J. D. Koralek} \affiliation{Materials Science
Division, Lawrence Berkeley National Laboratory, Berkeley,
California 94720, USA.}
\author{J. Orenstein\footnote{To whom all correspondence should be addressed. Email: jworenstein@lbl.gov}}
\affiliation{Department of Physics, University of California,
Berkeley, California 94720, USA.} \affiliation{Materials Science
Division, Lawrence Berkeley National Laboratory, Berkeley,
California 94720, USA.}
\author{D. R. Tibbetts}
\author{J. L. Reno}
\author{M. P. Lilly}
\affiliation{Sandia National Laboratories, Albuquerque, New Mexico
87123, USA.}


\begin{abstract}
We use phase-resolved transient grating spectroscopy to measure the
drift and diffusion of electron-hole density waves in a
semiconductor quantum well. The unique aspects of this optical
probe allow us to determine the frictional force between a
two-dimensional Fermi liquid of electrons and a dilute gas of holes. Knowledge of
electron-hole friction enables prediction of ambipolar dynamics in high-mobility
electron systems.
\end{abstract}
\maketitle

The motion of electrons and holes is crucial to the operation of
virtually all semiconductor devices and is a central topic of the
classic semiconductor texts \cite{Sze,Neamen}. In particular, the
coupled motion of electron-hole (\emph{e-h}) packets in applied
electric fields, known as ambipolar transport, is discussed in
depth.  However, it has been known for some time, although not
perhaps widely appreciated, that the motion of \emph{e-h} packets in
the high-mobility electron gases found in semiconductor quantum
wells and heterojunctions violates the predictions of the standard
theory. Insufficient understanding of ambipolar dynamics poses a
problem for the development of a spin-based electronics, as many
prospective devices are based on spin currents carried by spin
polarized \emph{e-h} packets subjected to electric fields
\cite{Zutic,Fabian,DDA}.

In the standard textbook description of ambipolar transport in a
doped semiconductor, electrons and holes interact only through the
long-range Coulomb interaction. Momentum relaxation occurs by
scattering on impurities and phonons and there is no exchange of
momentum between electrons and holes. On the basis of these
assumptions it is predicted that in an n-type semiconductor, for
example, an \emph{e-h} packet drifts in the direction of the force
on the holes, opposite to the motion of the Fermi sea of electrons.
However, by photoluminescence imaging, H\"{o}pfel \emph{et al.}
discovered that in GaAs quantum wells a drifting \emph{e-h} packet
moves in the direction of the majority, rather than minority
carrier, an effect they termed ``negative ambipolar mobility"
\cite{Hopfel1}. They recognized that this effect originates from the
scattering between electrons and holes, neglected in the standard
versions of ambipolar transport.

The scattering that dominates ambipolar transport in a single
quantum well is precisely analogous to the Coulomb drag effect that
has been studied intensively in systems in which layers of electron
gases are in close proximity \cite{Price, Gramila, Sivan}. In such
systems, the strength of the Coulomb interaction between layers can
be determined with precision via the transresistance, which is the
ratio of the voltage induced in one layer to a current in the other.
The transresistance is a direct measure of the rate of momentum
exchange (or frictional force) between the two coupled electronic
systems. Unfortunately, this technique cannot be used to probe the
much stronger frictional force between electrons and holes in the
same layer, which plays a crucial role in ambipolar dynamics.

In the experiments reported here we perform the first complete
characterization of coupled \emph{e-h} transport in a
two-dimensional electron gas (2DEG) by measuring simultaneously the
ambipolar diffusion coefficient $D_a$ and the ambipolar mobility
$\mu_a$.  From these measured coefficients, and a simple model of
momentum exchange between the Fermi sea and the packet, we obtain
the effective drag resistance $\rho_{eh}$ between electrons and
holes in a single quantum well. We show that the value of
$\rho_{eh}$ for a single layer, although orders of magnitude larger
the transresistance of bilayers, can be quantitatively understood
using the same random-phase approximation (RPA) model that describes
coupled quantum wells. Based on these findings, it becomes possible
to predict the ambipolar transport coefficients for high-mobility
semiconductors as a function of carrier density and temperature.

Our measurements of \emph{e-h} transport are performed using
transient grating spectroscopy (TGS) \cite{Eichler}, which is a
contact-free technique based on time-resolved optics. In TGS
standing waves of either \emph{e-h} or spin density \cite{Cameron}
are created in a 2DEG by photoexcitation with two noncollinear beams
of light from a pulsed (100 fs) laser. When the pulses are polarized
in the same direction, interference generates a standing wave of
laser intensity, creating a sinusoidal pattern of \emph{e-h} density
whose spatial period depends on the angle between the interfering
beams. The \emph{e-h} density wave imprinted in the 2DEG induces
local variation in the index of refraction, and therefore acts as an
optical diffraction grating. The time evolution of the density waves
after pulsed photogeneration can be monitored via the diffraction of
a time-delayed probe pulse.

The ambipolar diffusion coefficient can be readily determined by
measuring the rate at which the grating amplitude decays as a
function of its wavelength. However, as we discuss below,
characterization of $\rho_{eh}$ requires that $\mu_a$ must also be
measured under the same experimental conditions.  The latter is the
coefficient that relates the drift velocity of the \emph{e-h}
density wave to the magnitude of an electric field $E$ applied in
the plane of the 2DEG.  Measurement of $\mu_a$ clearly requires
sensitivity to the position of the \emph{e-h} density wave -
information that is contained in the \emph{phase shift} of the
diffracted light.  On the other hand, conventional scattering
experiments measure light intensity, and thus phase information is
lost. In the experiments reported here, we demonstrate that
time-resolved detection of both \emph{amplitude and phase} of light
diffracted from a drifting \emph{e-h} density wave allows
simultaneous determination $\mu_a$ and $D_a$, which together yield
the transresistance of the coupled \emph{e-h} system.

The measurements were performed on a 9 nm wide n-doped GaAs/AlGaAs
quantum well, grown by molecular beam epitaxy on a semi-insulating
GaAs (001) substrate (VB0355). The carrier density and mobility of
the 2DEG are $1.9\times10^{11}/\hbox{cm}^2$ and
$5.5\times10^{5}\hbox{cm}^2/\hbox{Vs}$ at 5 K, respectively. The
silicon donors were symmetrically doped in the center of each
barrier. The 2DEG channel was defined by a mesa etching, and Ohmic
contact was made by annealing NiGeAu to the sample. After patterning
the GaAs substrate was mechanically lapped and chemically etched to
allow for optical measurement in transmission geometry. Several
samples were prepared with semitransparent front and back gate
electrodes to allow for continuous variation of the equilibrium
electron density.

\begin{figure}
    \centering
    \subfigure[]{
      \label{fig:decayrate}
      \includegraphics[width=.46\textwidth]{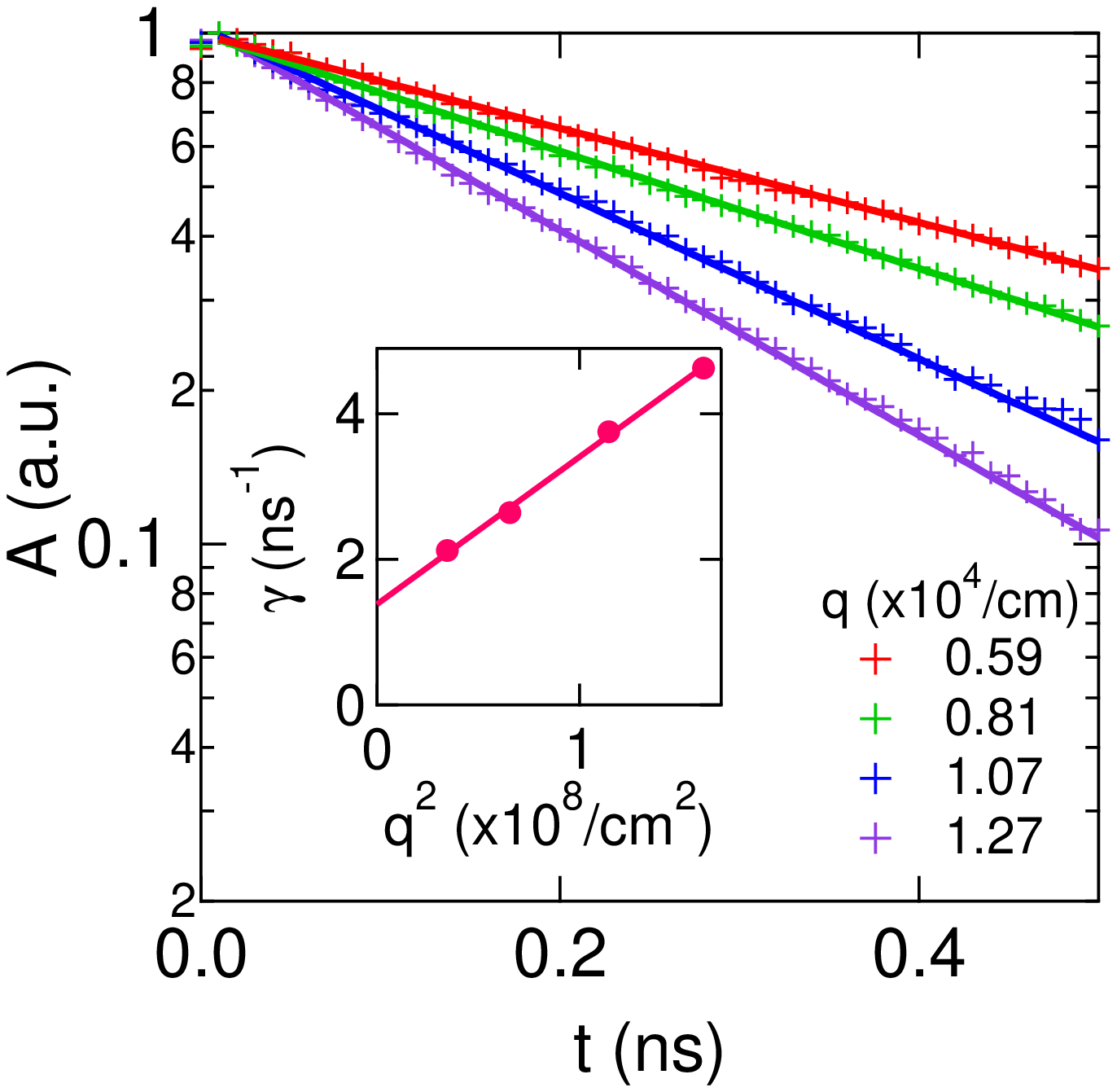}}
    \subfigure[]{
      \label{fig:phasechange}
              \includegraphics[width=0.46\textwidth]{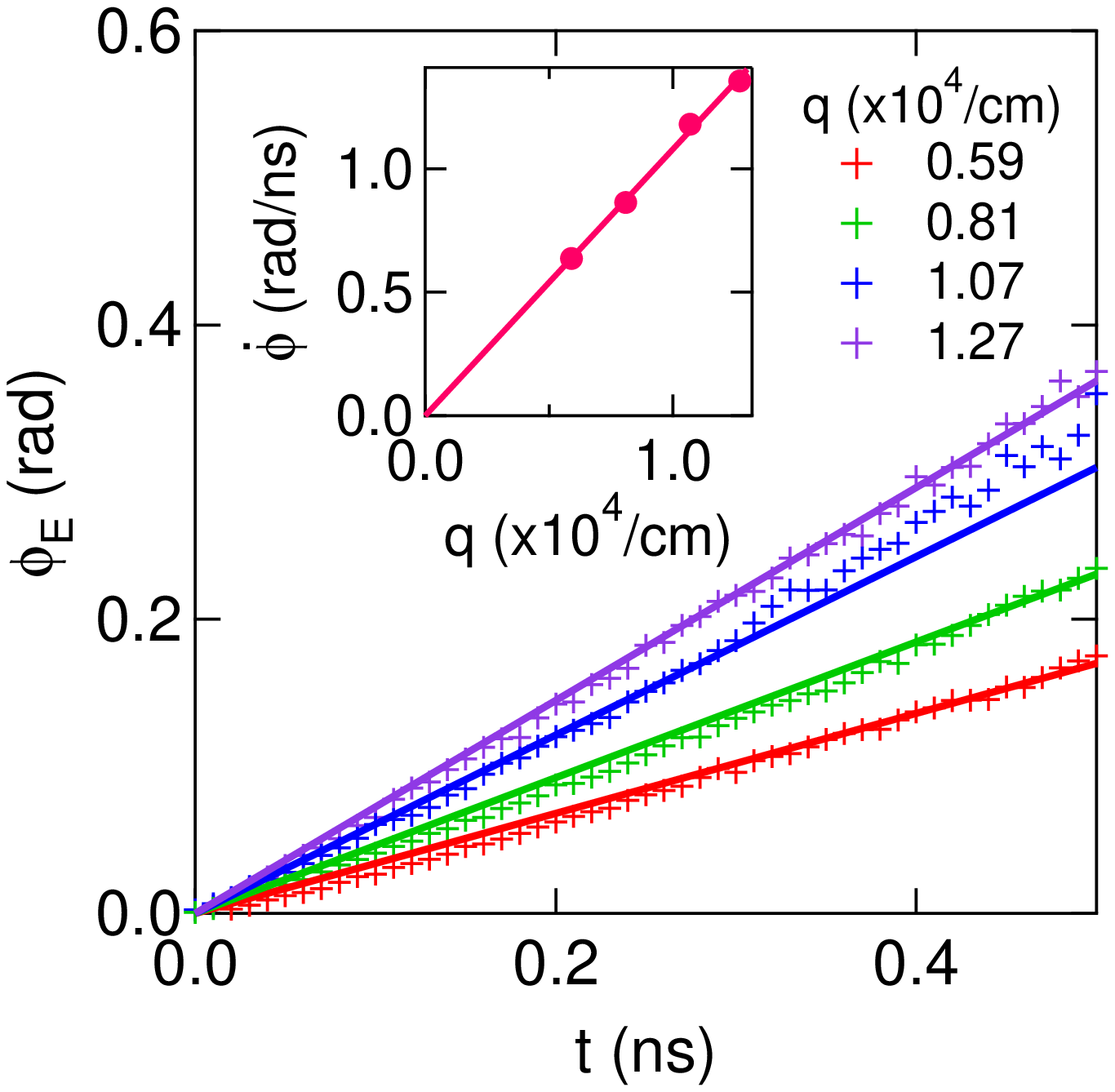}}
           \caption{(Color online) (a) Relative
amplitude of \emph{e-h} density wave as function of time for several
values of the wave vector $q$ measured at 50 K. Inset: The decay
rate $\gamma$ of the amplitude, plotted as a function of $q^2$; the
slope of the solid line through the data points is the ambipolar
diffusion coefficient $D_a$, and the intercept is the inverse of the
\emph{e-h} recombination time $\tau_{rec}$. (b) Linear advance of
the phase of the \emph{e-h} density wave with time for several
values of $q$, at 50 K. The applied electric field is
$E\approx2$V/cm. Inset: The rate of phase change $\dot{\phi}$ as a
function of $q$.
            }\label{fig:amplitude}
\end{figure}

The electron-hole density grating was generated by focusing the two
pump beams onto a 150 $\mu$m diameter spot between the two Ohmic
contacts, which are separated by 200 $\mu$m. Phase-sensitive
detection of the light diffracted from the grating was performed
using a heterodyne technique \cite{Vohringer, Chang, Goodno, Maznev,
Nuh1, Nuh2}, in which scattered pulses are mixed in a Si photodiode
with a beam of transmitted pulses acting as a local oscillator (LO).
The output voltage of the Si detector contains a phase-sensitive
term, proportional to $A(q,t)\exp[i(\phi_{pld}+\phi_{E})]$, where
$A(q,t)$ is the amplitude of the density wave, $\phi_{pld}=kd$
reflects the path length difference $d$ between the LO and
diffracted beams ($k$ is the wave vector of the light), and
$\phi_{E}=\mb{q}\cdot\delta\mb{r}$, where $\mb{q}$ and $\delta
\mb{r}$ are the grating wave vector and position, respectively. For
uniform motion with velocity $\mb{v}$ parallel to $\mb{q}$,
$\phi_E=qvt$. The linear advance of phase with time is equivalent to
a Doppler shift of frequency, $\Delta\omega=qv$. The phase noise
level of 0.01 rad in our detection system corresponds to an
uncertainty in velocity of $\sim10$ m/s, which is approximately 4
orders of magnitude smaller than the Fermi velocity $v_F$.

To measure $A(q,t)$ and $\phi_E(q,t)$ separately, we combine
heterodyne detection with two phase-modulation schemes.  To obtain
$A(q,t)$ we modulate $\phi_{pld}$ by oscillating the angle of a
coverslip placed in the LO beam path. For weak phase modulation the
synchronously detected heterodyne signal is proportional to
$A(q,t)\Delta\phi_{pld}$. To obtain $\phi_E(q,t)$ we oscillate the
in-plane $E$ field (applied parallel to $\mb{q}$) that induces drift
and measure the synchronous signal $A(q,t)v(E)qt$. From these two
measurements we extract $A(q,t)$ and $qv(E)t$ independently.

In Fig. \ref{fig:decayrate} we show the grating amplitude at a
representative temperature of 50 K as a function of time after
photogeneration, plotted on semilog axes, for several values of the
grating wave vector. The decay of $A(q,t)$ is a single exponential
with a rate constant, $\gamma$, that increases with increasing wave
vector.  As shown in the inset, $\gamma$ varies with $q$ as expected
for the combined effects of diffusion and electron-hole
recombination, $\gamma(q)=1/\tau_{rec}+D_aq^2$, where $\tau_{rec}$
is the electron-hole recombination time.

In Fig. \ref{fig:phasechange} we plot the phase of the \emph{e-h}
density wave, $\phi_E(q,t)$ versus $t$ for different values of $q$
at 50 K at full laser intensity $I_0 \simeq 0.25 \mu$J-cm$^{-2}$ per
pulse. The linear dependence of $\phi_E(q,t)$ on both $t$ and $q$
(see inset) is consistent with the Doppler shift
$\phi_E(q,t)=v(E)qt$. The sign of the phase shift gives the
direction of motion under the influence of the electric field, which
we determine to be the same as that of the electron Fermi sea. From
$\partial\phi_E(q,t)/\partial q$ at fixed time delay we obtain the
drift velocity of the \emph{e-h} density wave.  Normalizing by the
applied electric field yields the ambipolar mobility $\mu_a$.

\begin{figure}
    \centering
    \subfigure[]{
      \label{fig:mu_T}
      \includegraphics[width=0.46\textwidth]{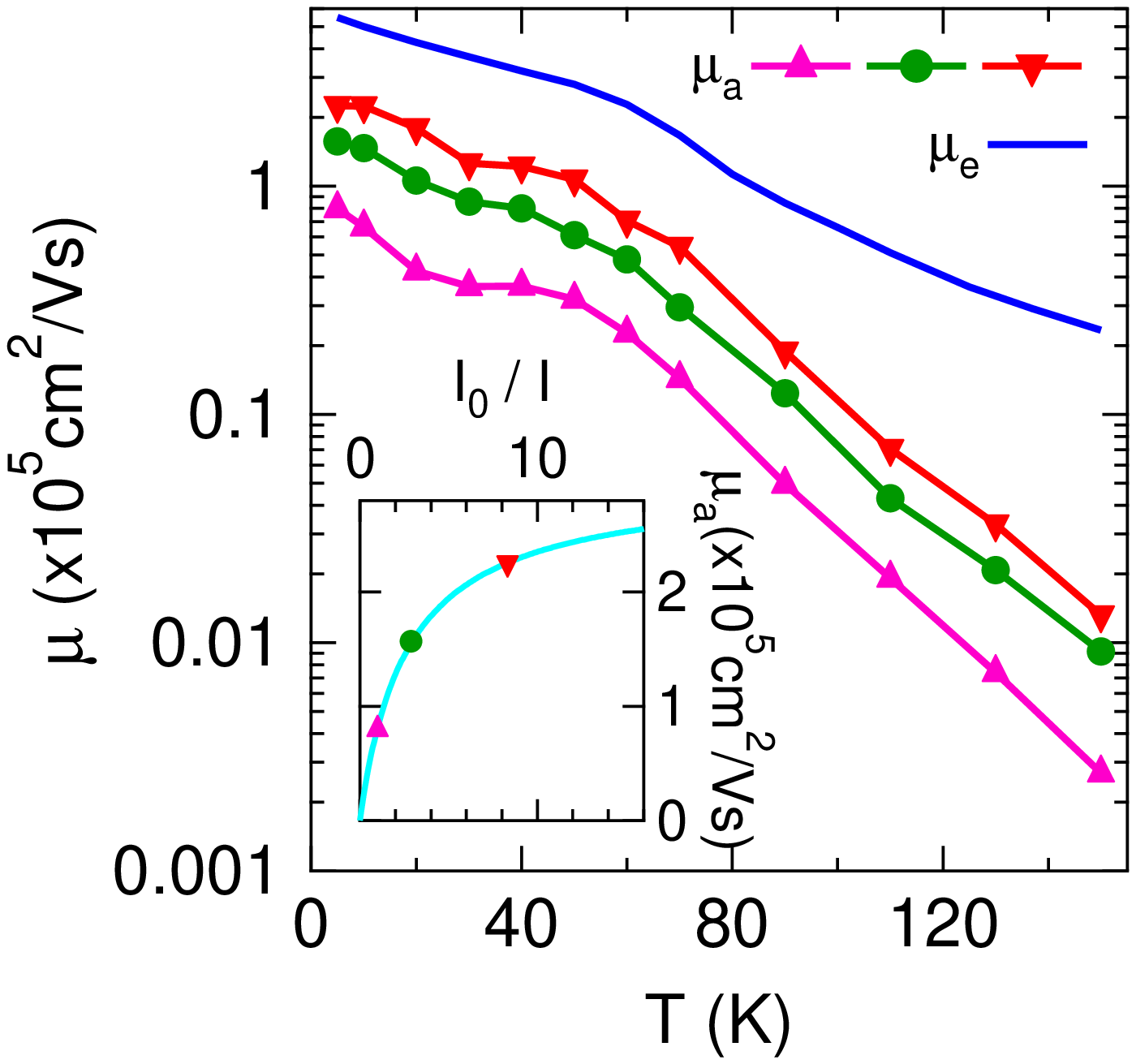}}
    \subfigure[]{
      \label{fig:gated}
      \includegraphics[width=0.46\textwidth]{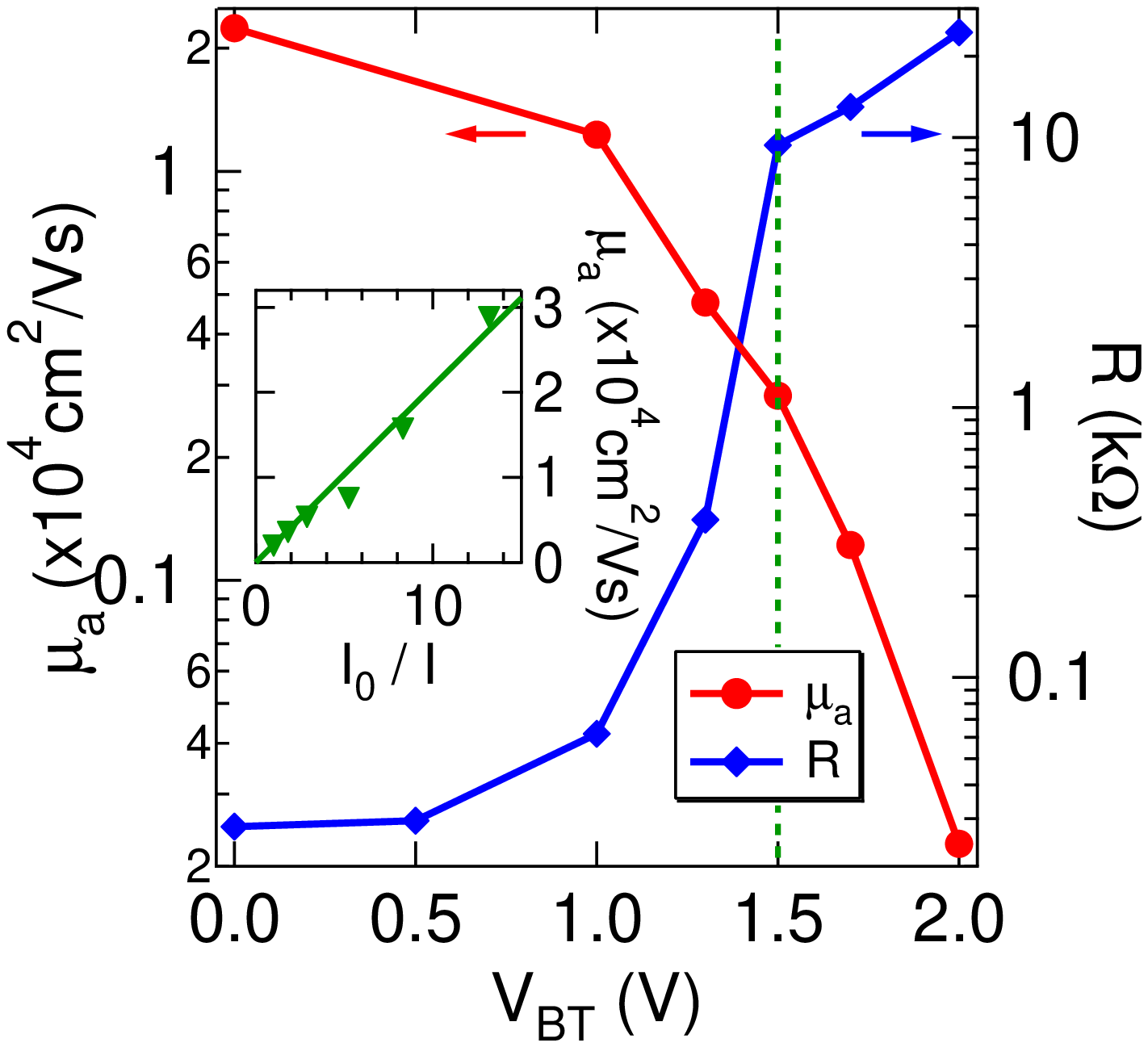}}
\caption{(Color online) (a) Ambipolar mobility $\mu_a$ at three
different pump intensities compared with electron mobility
 $\mu_e$, as a function of $T$. Inset: $\mu_a$ as a function of laser
intensity at 5 K, the solid line is a fit (see text). (b) Ambipolar
mobility $\mu_a$ and sample resistance $R$ as a function of gate
voltage $V_{BT}$ at 50 K. Inset: $\mu_a$ as a function of laser
intensity at fixed gate bias $V_{BT}$=1.5V at 50 K; solid line is a
linear fit showing that $\mu_a\propto 1/I$.}\label{fig:phase}
\end{figure}

In the course of the measurements we discovered that $\mu_a$ depends
strongly on $I$, in addition to the expected dependence on
temperature $T$. Figure \ref{fig:mu_T} shows $\mu_a$ determined
using the analysis outlined above as a function of $T$, for three
different values $I$.  For comparison, we also plot the electron
mobility $\mu_e$, as determined from standard four-contact dc
transport measurement.  As is clear from Fig. \ref{fig:mu_T},
$\mu_a$ decreases when either $T$ or $I$ increases.  When
nonequilibrium laser experiments show such dependencies, there can
be ambiguity as to whether the dependence on $I$ reflects an
intrinsic dependence on the photogenerated carrier density, $\Delta
n$, or the effect of transient local heating of the electron gas. To
determine whether the $I$ dependence is intrinsic, we performed TGS
measurements on a device with semitransparent gate electrodes, which
allowed us to vary the equilibrium electron density $n_0$ at fixed
$\Delta n$.

In Fig. \ref{fig:gated} we plot $\mu_a$ and the 2DEG resistance $R$
at 50 K, as a function of the voltage between the two gates,
$V_{BT}$. Clearly $\mu_a$ decreases rapidly as $n_0$ is driven to
zero (and $R\rightarrow\infty$) by increasingly positive $V_{BT}$.
As these measurements are performed at constant $I$, it is evident
that the intensity dependence shown in Fig. \ref{fig:mu_T} reflects
an intrinsic dependence of $\mu_a$ on the ratio $\Delta n/n_0$,
rather than laser-induced heating.  The inset of Fig.
\ref{fig:gated} illustrates that $\mu_a$ scales as $1/I$ (equivalent
to $1/\Delta n$) in the regime where $n_0$ is small, while the Fig.
\ref{fig:mu_T} inset shows that $\mu_a$ approaches an asymptotic
value $\mu_{a0}$ in the limit that $I$ (and $\Delta n$) $\rightarrow
0$. The overall dependence of variation of $\mu_a$ can be summarized
by the simple formula

\begin{equation}\label{eq:mua_fit}
\mu_a(I)=\frac{\mu_{a0}}{1+\alpha (\Delta n/n_0)},
\end{equation}
where $\alpha$ is a $T$-dependent parameter.

\begin{figure}
\centering
    \subfigure[]{
      \label{fig:mu_0}
      \includegraphics[width=0.46\textwidth]{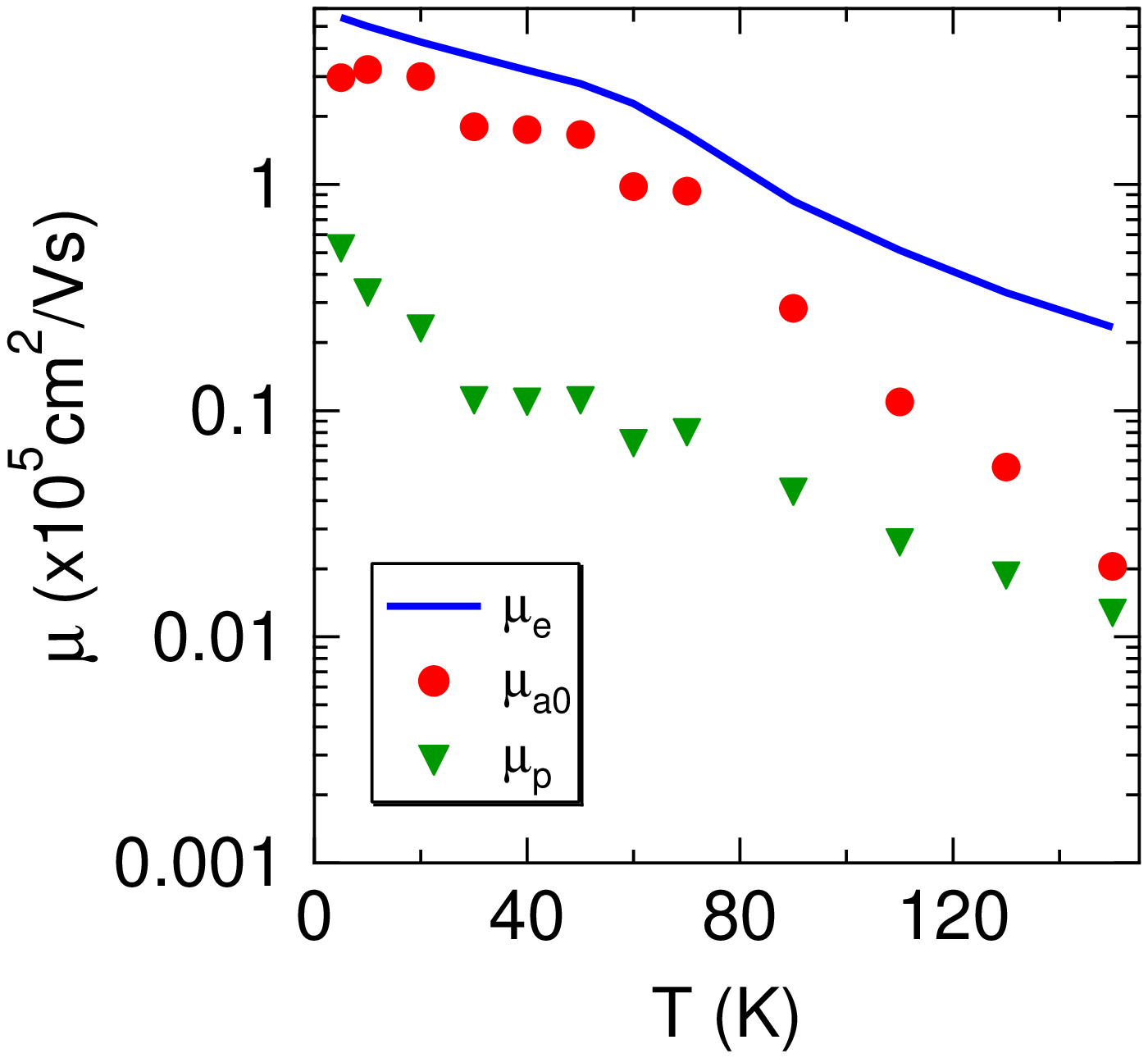}}
    \subfigure[]{
      \label{fig:Einstein_rho_eh}
      \includegraphics[width=0.46\textwidth]{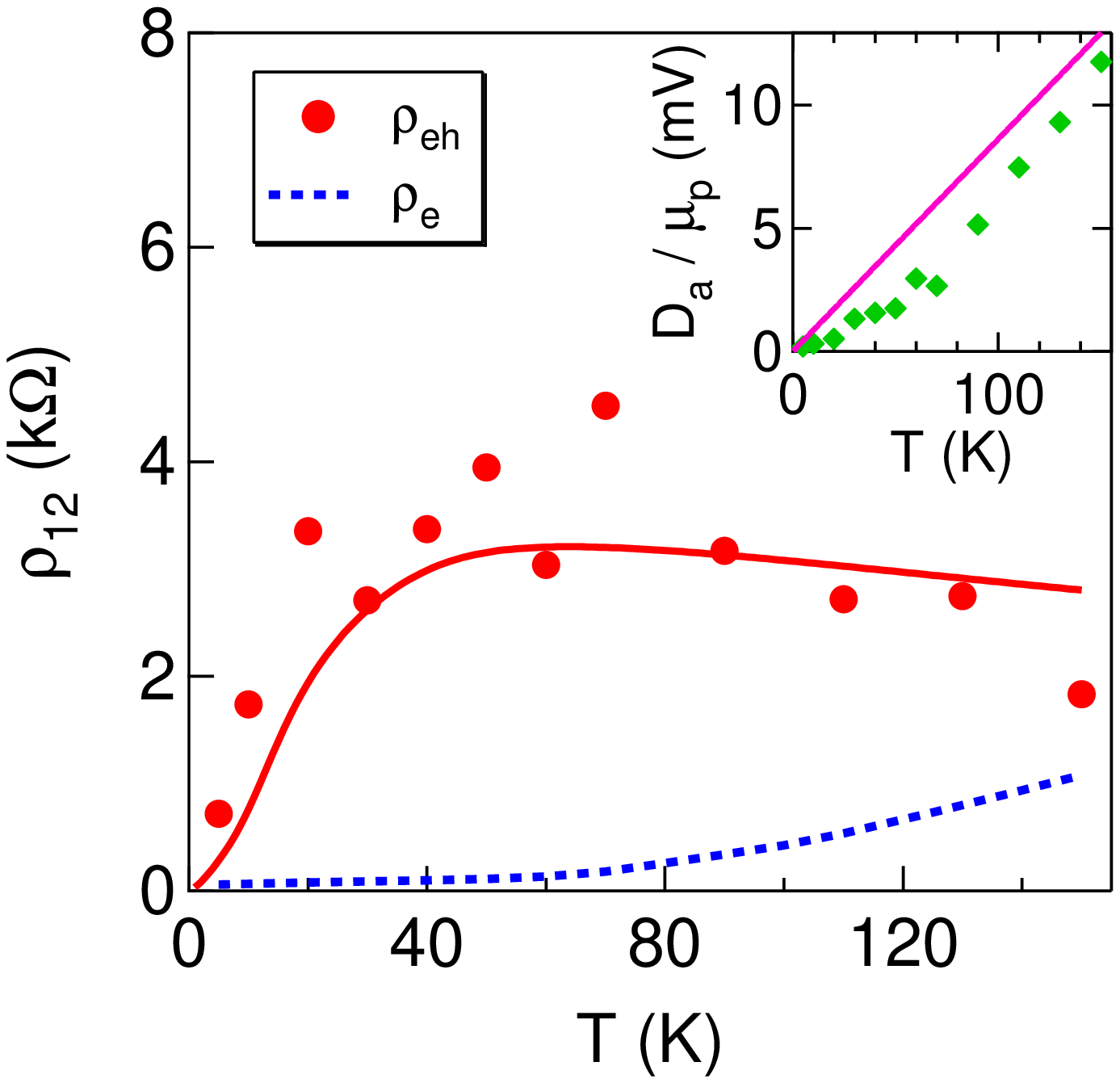}}
\caption{(Color online) (a) Comparison of electron mobility $\mu_e$;
ambipolar mobility $\mu_{a0}$, and packet mobility $\mu_p$, as a
function of $T$. (b) The \emph{e-h} drag transresistivity
$\rho_{eh}$ as a function of $T$; solid line is a theoretical
prediction of $\rho_{eh}$ based on the RPA. Inset: $D_a/\mu_p$
compared with the Einstein relation prediction in the non-degenerate
regime $k_BT/e$. }
\end{figure}

At this point, we can summarize our experimental findings as
follows: (1) The photogenerated \emph{e-h} packet drifts under the
influence of an $E$ field in the same direction as the Fermi sea of
electrons, (2) the velocity of the packet goes to zero as $\Delta
n/n_0 \rightarrow \infty$ and approaches a constant in the limit
that $\Delta n/n_0 \rightarrow 0$, (3) the asymptotic value,
$\mu_{a0} (T)$ [Fig. \ref{fig:mu_0}], is proportional to, but
slightly smaller than, the electron mobility for $T<80$ K, but
becomes much smaller than $\mu_e$ for $T>80$ K.  We show below that
each of these observations can be understood with a relatively
simple model that treats the \emph{e-h} packet as a neutral gas of
particles that can exchange momentum with the Fermi sea.

The stationary transport equations for free electrons and the packet
can be written as
\[\begin{split}
&\frac{n_0m_ev_e}{\tau_{e}}+n_0\Delta n\gamma(v_e-v_p)=-n_0eE,\\
&\frac{\Delta nm_pv_p}{\tau_{p}}+n_0\Delta
n\gamma(v_p-v_e)+k_BT\nabla (\Delta n)=0,
\end{split}\] where $1/\tau_{e(p)}$ is the rate at which electrons(packet) lose momentum to the lattice, $m_e$ and $m_p$ are respective masses, and
$\gamma$ is a parameter describing the rate of momentum exchange. By
solving these equations we obtain precisely the form of Eq.
\ref{eq:mua_fit}, where
\begin{equation}\label{eq:mu_a0}
\mu_{a0}=-\frac{\mu_e}{1+\dfrac{\mu_{e}}{\mu_p}\dfrac{\rho_e}{\rho_{eh}}},
\end{equation}
and $\alpha=\mu_{a0}/\mu_p$.  In Eq. \ref{eq:mu_a0} we have made use
of the definitions, $\mu_p\equiv e\tau_{p}/m_p$, $\rho_e\equiv (n_0
e \mu_e)^{-1}$, and $\gamma\equiv e^2 \rho_{eh}$. The negative sign
of $\mu_a$ corresponds to the \emph{e-h} packet drifting in the same
direction as the Fermi sea of electrons. In addition, we find that
solving for the ambipolar diffusion coefficient yields
\begin{equation}\label{eq:D_a}
D_a=\frac{k_BT\mu_{a0}}{e}\frac{\rho_e}{\rho_{eh}}.
\end{equation}

From Eq. \ref{eq:D_a} we see that independent measurement of
$\mu_{a0}$, $D_a$, and $\rho_e$ directly yields the electron-hole
transresistance, $\rho_{eh}$. The values of $\rho_{eh}$ thus
determined are plotted versus $T$ in Fig. \ref{fig:Einstein_rho_eh},
together with $\rho_e$ for comparison. We see that in the low $T$
regime, $\rho_e \ll \rho_{eh}$, which translates to an ambipolar
mobility that is not too different from the electron mobility. As
$T$ increases and $\rho_e$ approaches $\rho_{eh}$, $\mu_{a0}$ tends
towards the much smaller $\mu_p$. While the values of ambipolar
mobility are controlled by $\rho_e/\rho_{eh}$, $D_a$ itself is
fairly insensitive to Coulomb drag because diffusive spreading of
the packet takes place with parallel transport of electrons and
holes. This effect is illustrated in the inset of Fig.
\ref{fig:Einstein_rho_eh}, which compares the ratio $D_a/\mu_p$ to
$k_BT/e$.  The near agreement with the Einstein relation shows that
$D_a$ is essentially determined by the nondegenerate gas of holes
because the electrons in the packet are tethered to them through the
long-range Coulomb interaction.

The values of $\rho_{eh}(T)$ that we obtain are several orders of
magnitude larger than those obtained in Coulomb drag experiments on
coupled quantum wells \cite{Croxall, Lilly}. However, in the
experiments reported here (\emph{i}) electrons and holes are
confined to the same quantum well and (\emph{ii}) one of the Fermi
gases (the holes) is nondegenerate throughout the $T$ range of the
experiment. To test whether the values of $\rho_{eh}(T)$ shown in
Fig. \ref{fig:Einstein_rho_eh} are reasonable, we apply the standard
RPA model for Coulomb drag to the single layer case. The RPA
expression for $\rho_{eh}(T)$ is the phase space integral of the
product of the interaction, $V_{\text{RPA}}(q)$, and
Im$\{\chi_{1,2}\}$, the imaginary part of the susceptibility of
fermion species 1 and 2, respectively
\cite{Flensberg,Zheng,Vignale}.  To apply this theory to our
experiment, we substitute the nondegenerate Lindhard response for
the hole susceptibility \cite{Vignalebook}. Numerical evaluation of
the phase space integral, plotted as a solid line in Fig.
\ref{fig:Einstein_rho_eh}, shows that the RPA interaction describes
the experimental data quite well without any free parameters.

In conclusion, we have used phase-resolved TGS to simultaneously
measure the ambipolar drift and diffusion of photoinjected electrons
and holes. From these measurements we determine for the first time
the frictional force between a degenerate Fermi liquid of electrons
and a dilute gas of holes in the same two-dimensional system. The
measured values of $\rho_{eh}$ data are accurately described by the
static limit of RPA-based theory with no free parameters. By
combining $\rho_{eh}$ with a simple model based on conservation of
momentum, the ambipolar dynamics of high-mobility electron gases can
be predicted, enabling more powerful modeling of devices, for
example those based on spin current of drifting polarized carriers.

\begin{acknowledgments}
All the optical and some of the electrical measurements were carried
out at Lawrence Berkeley National Laboratory and were supported by
the Director, Office of Science, Office of Basic Energy Sciences,
Materials Sciences and Engineering Division, of the U.S. Department
of Energy under Contract No. DE-AC02-05CH11231. Sample growth and
processing and some of the transport measurements were performed at
the Center for Integrated Nanotechnologies, a U.S. Department of
Energy, Office of Basic Energy Sciences user facility at Sandia
National Laboratories (Contract No. DE-AC04-94AL85000).
\end{acknowledgments}


\begin{thebibliography}{}


\bibitem{Sze} S.M. Sze and K.K. Ng,
\emph{The Physics of Semiconductor Devices}, (John Wiley and Sons, New
York, 2003).

\bibitem{Neamen} D.A. Neamen, \emph{Semiconductor Physics and
Devices: basic principles} 3rd ed. Ch. 6 (McGraw-Hill Higher
Education, Boston, 2003).

\bibitem{Zutic} I. Zutic, J. Fabian, and S. Das Sarma, Rev. Mod. Phys. \textbf{76}, 323 (2004).

\bibitem{Fabian} J. Fabian, A. Matos-Abiague, C. Ertler, P. Stano,
and I. \v{Z}uti\'{c}, Acta Phys. Slov. \textbf{57}, 565 (2007).

\bibitem{DDA} D.D. Awschalom and M.E. Flatt\'{e}, Nature Phys. \textbf{3}, 153 (2007).


\bibitem{Hopfel1} R.A. H\"{o}pfel, J. Shah, P.A. Wolff, and A.C. Gossard, Phys. Rev. Lett. \textbf{56}, 2736 (1986).

\bibitem{Price} P.J. Price, Physica (Amsterdam) \textbf{117B+C}, 750 (1983).

\bibitem{Gramila} T.J. Gramila, J.P. Eisenstein, A.H. MacDonald, L.N. Pfeiffer, and K.W. West, Phys. Rev. Lett. \textbf{66}, 1216 (1991).

\bibitem{Sivan} U. Sivan, P.M. Solomon, and H. Shtrikman, Phys. Rev. Lett. \textbf{68}, 1196 (1992).

\bibitem{Eichler} H.J. Eichler, P. Gunter, and D.W. Pohl, \emph{Laser-Induced
Dynamic Gratings} (Springer-Verlag, Berlin, 1986).

\bibitem{Cameron} A.R. Cameron, P. Riblet, and A. Miller, Phys. Rev. Lett. \textbf{76}, 4793 (1996).

\bibitem{Vohringer} P. Vohringer and N.F. Scherer, J. Phys. Chem. \textbf{99}, 2684 (1995).

\bibitem{Chang} Y.J. Chang, P. Cong, and J.D. Simon, J. Phys. Chem. \textbf{99}, 7857
(1995).

\bibitem{Goodno} G.D. Goodno, G. Dadusc, and R.J.D. Miller, J. Opt. Soc. Am. B \textbf{15}, 1791 (1998).

\bibitem{Maznev} A.A. Maznev, K.A. Nelson, and J.A. Rogers, Opt. Lett. \textbf{23}, 1319 (1998).

\bibitem{Nuh1} N. Gedik and J. Orenstein, Opt. Lett. \textbf{29}, 2109 (2004).

\bibitem{Nuh2} N. Gedik, J. Orenstein, R. Liang, D.A. Bonn, and W.N. Hardy, Science \textbf{300}, 1410 (2003).

\bibitem{Croxall} A.F. Croxall \emph{et al.}, Phys. Rev. Lett. \textbf{101}, 246801 (2008).

\bibitem{Lilly} J.A. Seamons, C.P. Morath, J.L. Reno, and M.P. Lilly, Phys. Rev. Lett. \textbf{102}, 026804 (2009).

\bibitem{Zheng} L. Zheng and A.H. MacDonald, Phys. Rev. B \textbf{48}, 8203 (1993).

\bibitem{Flensberg} K. Flensberg and Ben Yu.-Kuang. Hu, Phys. Rev. B  \textbf{52}, 14796 (1995).

\bibitem{Vignale} G. Vignale and A.H MacDonald, Phys. Rev. Lett. \textbf{76}, 2786 (1996).

\bibitem{Vignalebook} G.F. Giuliani and G. Vignale,
\emph{Quantum theory of the electron liquid} Ch. 4 (University
Press, Chambridge, 2005).

\end{thebibliography}
\end{document}